\documentstyle[aps,epsf]{revtex} 
\begin{document} 
\widetext
 
\title{ Renormalization group approach to the interacting \\
        boson fermion systems}

\author{T.~Doma\'nski$^{(1)}$ and J.~Ranninger$^{(2)}$} 
\address{\sl $^{(1)}$ Institute of Physics, M.~Curie Sk\l odowska 
             University, 20-031 Lublin, Poland \\
$^{(2)}$ Centre de Recherches sur les Tr\`es Basses Temperatures,
         CNRS, 38-042 Grenoble, France}

\date{\today}  
\maketitle 
\draft 

\begin{abstract}
We study a pseudogap region of the mixed boson fermion system 
using a recent formulation of the renormalization group technique 
through the set of infinitesimal unitary transformations.
Renormalization of fermion energies gives rise to a depletion of 
the low energy states (pseudogap) for temperatures $T^{*}>T>T_{c}$ 
which foreshadows appearance of the pairwise correlations with 
their long range phase coherence being missed. With a help of 
the flow equations for boson and fermion operators we analyze 
spectral weights and finite life times of these quasiparticles 
caused by interactions.
\end{abstract}

\vspace{2cm}
\noindent
Many experimental measurements of e.g.\ the angle resolved
photoemission (ARPES), tunneling spectroscopy (STM), specific 
heat as well as the optical and magnetic characteristics 
(see for an overview \cite{Timusk}) point at a presence of 
pseudogap in the normal phase spectrum of underdoped cuprate
superconductors. It is believed that a proper description 
of this phenomenon would be crucial for understanding 
of the HTSC mechanism.

Model of itinerant fermions coupled via the charge
exchange interaction to boson particles \cite{BFM}, 
for example of bipolaronic origin, has been shown 
to posses such a pseudogap in the normal phase fermion 
spectrum. Its occurrence has been obtained on a basis
of the selfconsistent perturbation theory \cite{Robin},
diagrammatic expansion \cite{Ren} and also by solving 
the dynamical mean field equations for this model
within the non-crossing approximation (NCA) \cite{NCA}. 
Unfortunately, these techniques have not been able 
to study a superconducting phase of the model, so
evolution of the pseudogap into the true superconducting
gap could not be investigated. Only very recently 
we have reported some prior results for the boson fermion
(BF) model using the flow equation method \cite{Domanski} 
where superconducting gap and normal phase pseudogap 
could be treated on equal footing. Both gaps originate 
from the same source (from the exchange interaction 
between fermions and bosons) but they are differently 
scaled against a magnitude of this interaction and 
the total concentration of charge carriers. Still 
the most intriguing question concerning the evolution 
from the pseudo- to the superconducting gap is poorly
understood due to technical difficulties (it is a very
tough problem to solve the flow equations for $2+
\epsilon$ dimensional system). Some attempts in
this direction are currently under consideration.

In this paper we address another important issue
concerning life time effects of the quasiparticles. 
The high resolution ARPES data \cite{Campuzano} 
reveal that normal phase of underdoped cuprates 
has the marginal Fermi liquid type properties 
where the quasiparticle weight and life time 
are very sensitive to temperature. We want to check 
how the life times of fermions and bosons depend 
on temperature $T$ and momentum using the BF model 
scenario.

Flow equation for any arbitrary operator, say $\hat{A}(l)$,
is given by
\begin{eqnarray}
\frac{d \hat{A}(l)}{dl} = \left[ \hat{\eta}(l),\hat{A}(l) \right]
\label{flow}
\end{eqnarray}
where $\hat{\eta}$ is a generating operator of the continuous
canonical transformation. The initial BF model Hamiltonian
consists of the free part
\begin{eqnarray}
\hat{H}_{0} = \sum_{k,\sigma} (\varepsilon_{k}^{\sigma} - \mu)
        \hat{c}_{k\sigma}^{\dagger}\hat{c}_{k\sigma} +
        \sum_{q} (E_{q}-2\mu) \hat{b}_{q}^{\dagger}\hat{b}_{q} 
\end{eqnarray}
and the interaction 
\begin{eqnarray}
\hat{H}_{int} = \frac{1}{\sqrt N} \sum_{k,p} \left( v_{k,p} 
\hat{b}^{\dagger}_{k+p} \hat{c}_{k\downarrow}\hat{c}_{p\uparrow} 
+ v^{*}_{k,p} \hat{b}_{k+p} \hat{c}^{\dagger}_{p\uparrow}
\hat{c}^{\dagger}_{k\downarrow} \right).
\end{eqnarray}
In order to decouple fermion from boson subsystems
we have previously chosen $\eta$ in a form \cite{Domanski}
\begin{eqnarray}
\hat{\eta}(l) = - \sum_{k,q} \left( \alpha_{k,q}(l) v_{k,q}(l)
\hat{b}^{\dagger}(l) \hat{c}_{k,\downarrow}(l) 
\hat{c}_{q,\uparrow}(l) - h.c. \right)
\label{eta}
\end{eqnarray} 
with $\alpha_{k,q}=\varepsilon_{k}^{\downarrow} +
\varepsilon_{q}^{\uparrow} - E_{k+q}$. Using the general flow
equation (\ref{flow}) we transformed the Hamilton operator 
$\hat{H}(l)$ so that in the limit $l \rightarrow \infty$
the charge exchange interaction got eliminated $v_{k,q}
(\infty)=0$.

In a course of the continuous canonical transformation 
the initial boson and fermion operators change according
to the following flow equations
\begin{eqnarray}
\frac{d \hat{b}_{q}(l)}{dl} & = & \sum_{k} \alpha_{k,q-k}(l)
v_{k,q-k}(l) \hat{c}_{k,\downarrow}(l) \hat{c}_{q-k,\uparrow}(l)
\label{b_flow} \\
\frac{d \hat{c}_{k,\uparrow}(l)}{dl} & = & - \sum_{q}
\alpha_{q,k}(l) v_{k,q}(l) \hat{b}_{q+k}(l) 
\hat{c}_{q,\downarrow}^{\dagger}(l).
\label{c_flow}
\end{eqnarray}
Similar equations can be derived for their hermitean conjugates 
$\hat{b}_{q}^{\dagger}(l)$, $\hat{c}_{k,\uparrow}^{\dagger}(l)$. 
By inspecting (\ref{b_flow}) we notice that initial boson 
annihilation operator is coupled to the electron pair operator 
$\hat{c}_{k,\downarrow}(l) \hat{c}_{q-k,\uparrow}(l)$.
Thereof one should in a next step determine a corresponding 
flow equation for the finite momentum pair operator, however 
this process will never end (like equations of motion for 
the Greens functions). There would be the higher and higher 
order product of operators getting involved into the differential 
flow equations. 

Following the other works using such continuous canonical 
transformation \cite{spin-boson}-\cite{Hofstetter} we 
terminate the flow equations by postulating some (physical) 
Ansatz. In a context of the BF model it seems natural 
to postulate the following decompositions
\begin{eqnarray}
\hat{b}_{q}(l) & = & X_{q}(l) \hat{b}_{q} + \sum_{k} Y_{q,k}(l)
\hat{c}_{k,\downarrow} \hat{c}_{q-k,\uparrow}
\label{b_ansatz}, \\
\hat{c}_{k,\sigma}(l) & = & A_{k}(l) \hat{c}_{k,\sigma} +
\sum_{q} B_{k,q}(l) \hat{b}_{q+k} \hat{c}_{q,-\sigma}^{\dagger}
\label{c_ansatz},
\end{eqnarray}
where $\hat{b}_{q} \equiv \hat{b}_{q}(l=0)$ and 
$\hat{c}_{k,\sigma} \equiv \hat{c}_{k,\sigma}(l=0)$.
The initial $l=0$ values are 
\begin{eqnarray}
\begin{array}{ll} 
A_{k}(0) = 1, & \hspace{1cm} B_{k,q}(0)  =  0, \\
X_{q}(0) = 1, & \hspace{1cm} Y_{k,q}(0) = 0.
\end{array}
\end{eqnarray}

Unknown parameters $Y_{q,k}(l)$, $B_{k,q}(l)$ can be 
found directly from the flow equations (\ref{b_flow},\ref{c_flow})
\begin{eqnarray}
\frac{dB_{k,q}(l)}{dl} & = & - \alpha_{q,k}(l) 
v_{q,k}(l) X_{q+k}(l)A_{q}(l) \label{flow1} \\
\frac{dY_{q,k}(l)}{dl} & = & \alpha_{k,q-k}(l) 
v_{k,q-k}(l) A_{k}(l) A_{q-k}(l). \label{flow2}
\end{eqnarray}
There is no unambiguous way for determining the other
missing coefficients $A_{k}(l)$, $X_{q}(l)$ because with
the Ansatz (\ref{b_ansatz},\ref{c_ansatz}) the flow 
equations still do not close. In what follows bellow 
we determine these parameters using the constraint
\begin{eqnarray}
\left[ b_{q}(l), b_{q'}^{\dagger}(l) \right] 
& = & \delta_{q,q'}, 
\label{commut} \\
\left\{ c_{k,\sigma}(l),c_{k',\sigma'}^{\dagger}(l) \right\} 
& = & \delta_{k,k'} \delta_{\sigma, \sigma'}.
\label{anti}
\end{eqnarray}
These statistical commutation/anticommutation relations yield
\begin{eqnarray}
A_{k}(l)^{2} + \sum_{q} B_{k,q}(l)^{2} \left( 
f_{q,\downarrow} - b_{k+q} \right) & = & 1, \label{flow3}\\
X_{q}(l)^{2} + \sum_{k} Y_{q,k}(l)^{2} \left( 1 -
f_{k,\downarrow} - f_{q-k,\uparrow} \right) & = & 1. 
\label{flow4}
\end{eqnarray}
Here $f_{k,\sigma}$, $b_{q}$ denote the Fermi and Bose 
distribution functions. They appear after introducing
the normal ordered forms for operators left of the
commutation (\ref{commut}) and anticommutation (\ref{anti})
when substituting (\ref{b_ansatz},\ref{c_ansatz}).

We solved numerically a set of the coupled equations 
(\ref{flow1},\ref{flow2},\ref{flow3},\ref{flow4})
simultaneously with the additional flow equations 
for $\varepsilon_{k}^{\sigma}(l)$,$E_{k}(l)$ and 
$v_{k,q}(l)$ given in the Ref.\ \cite{Domanski}. 
Initially, for $l=0$, we assumed: a) the one dimensional 
tight binding dispersion $\varepsilon_{k}^{\sigma}
(l=0)=-2t\cos{k}$, b) localized bosons $E_{q}(l=0)=
\Delta_{B}$ and c) local exchange interaction 
$v_{k,q}(l=0)=v$. For the sake of comparison with 
the previous studies of this model we used $D=4t\equiv1$,
$\Delta_{B}=-0.6$, $v=0.1$ and total concentration of 
carries fixed to be 1. 

\begin{figure}
\centerline{\epsfxsize=8cm \epsfbox{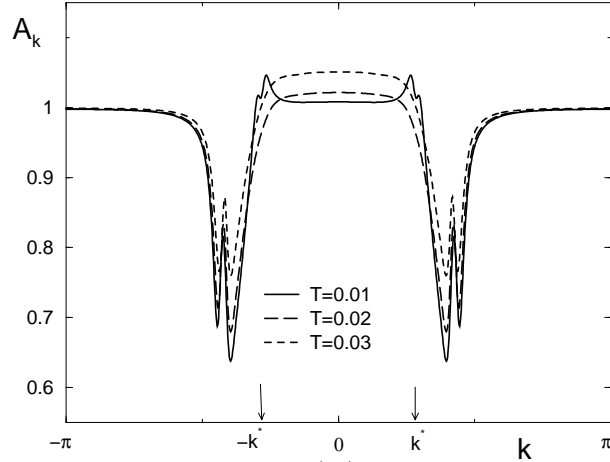}}
\caption{Momentum dependence of the fermion
coefficient $A_{k}(\infty)$ for three
temperatures $T=0.01$, $0.02$ and $0.03$
(in units of the initial fermions bandwidth).
Note a strong renormalization of the coefficient
$A_{k}$ around the characteristic momentum $k^{*}$.}
\end{figure}

In figure 1 we plot the momentum dependence of $A_{k}$ 
coefficient for $l=\infty$ (when fermion subsystem is 
decoupled from the boson one). Considerable variation 
of the parameter $A_{k}$ from its initial value 
(being $1$) can be observed for such momenta which 
are located near $k^{*}\simeq 0.89$. As discussed
in the Ref.\ \cite{Domanski} this situation corresponds
to the resonant scattering processes when $\varepsilon_{k*}
+\varepsilon_{-k*}=\Delta_{B}$. It is worth mentioning
that for $n_{tot}=1$ the Fermi momentum approaches 
this $k^{*}$ from bellow when temperature decreases.
We could think of the coefficient $A_{k}$ as a
measure of the spectral weight for fermions. The 
missing part $| 1 - A_{k}| $ is transfered towards
some composite objects (mixtures of fermions and 
bosons or fermion pairs). Taking into account 
the fact that {\em for a decreasing temperature $k_{F}$ 
shifts closer and closer to $k^{*}$, we shall thus 
observe a diminishing spectral weight of fermions}. 
This is in agreement with interpretation of the ARPES 
experimental results \cite{Campuzano}. 

\begin{figure}
\centerline{\epsfxsize=8cm \epsfbox{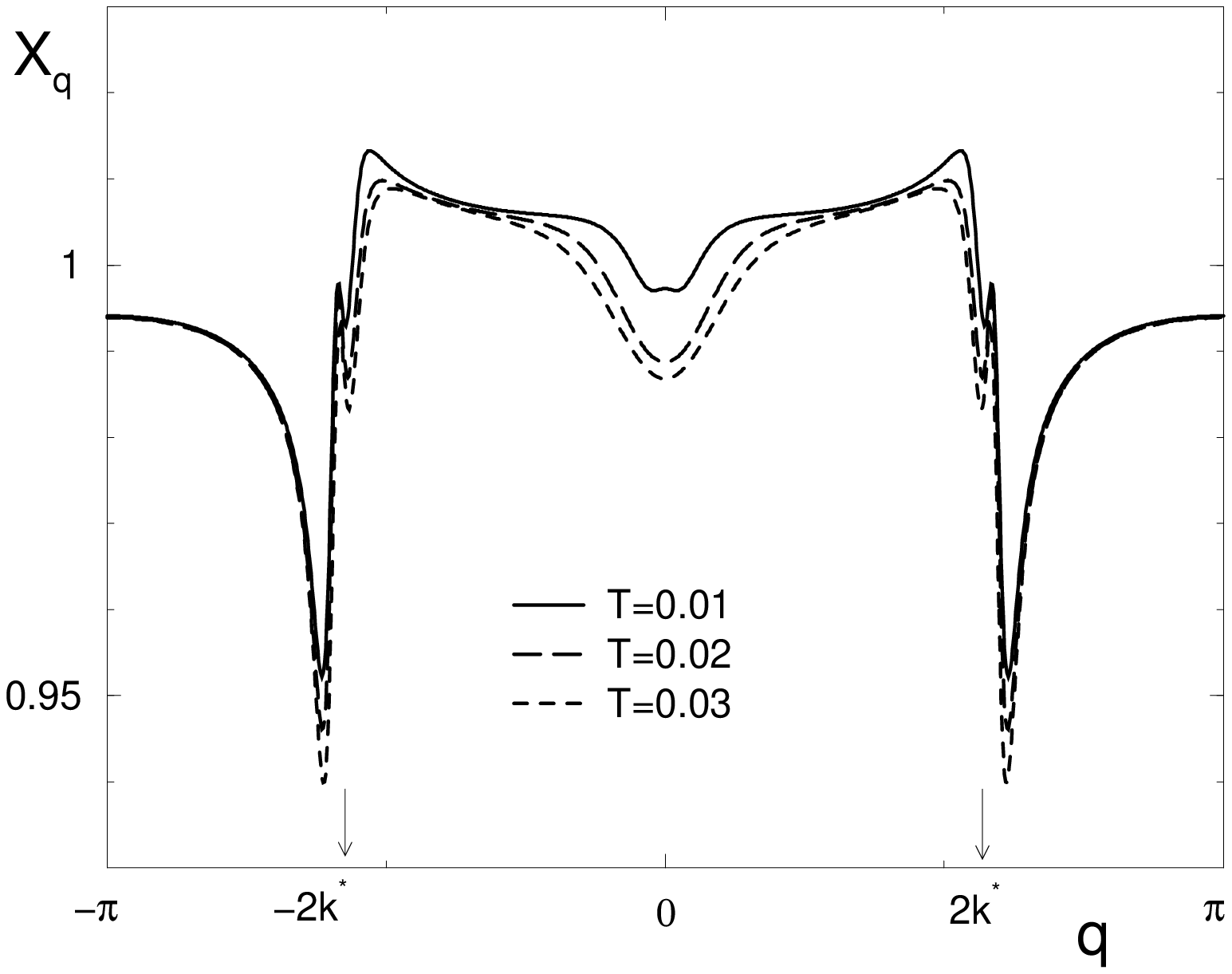}}
\centerline{\epsfxsize=8cm \epsfbox{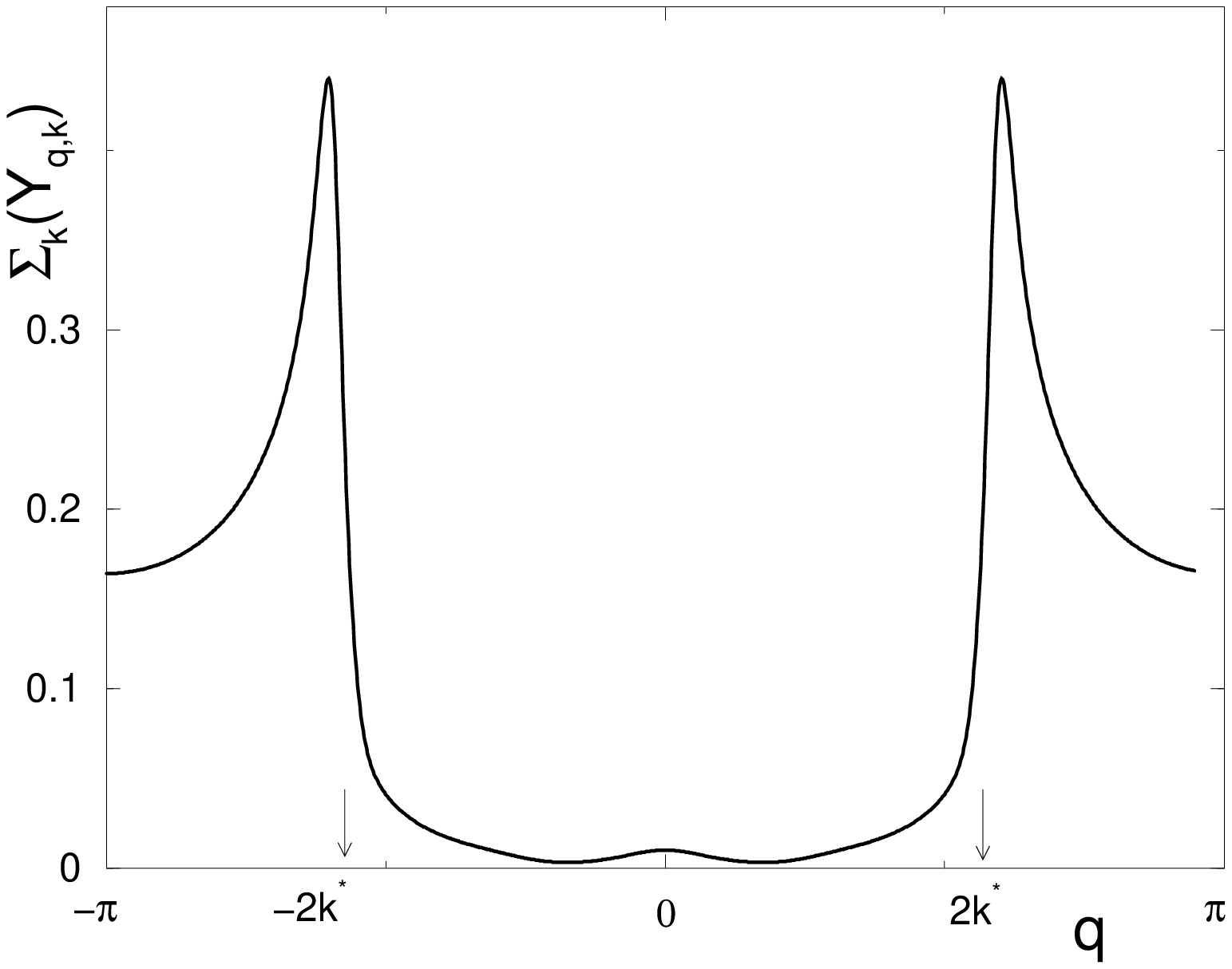}}
\caption{Momentum dependence of the boson
coefficients $X_{q}(\infty)$ and $\sum_{k}Y_{q,k}
(\infty)$ for the same temperatures as in figure 1.
The most efficient renormalization of boson 
coefficients occurs near $q \sim 2k^{*}$.}
\end{figure}

For a completeness we illustrate also the resulting 
($l=\infty$) values of the boson coefficients given in
(\ref{b_ansatz}). Parameter $X_{q}$ shows only a small
deviation from its initial value $1$. This time such
a change is most effective for boson momenta close to
$2k^{*}$. It is exactly the same region of the Brillouin 
zone where the effective boson dispersion shows a kink
(see figure 3). 

The missing part of the boson spectral weight 
is transferred to the fermion pairs, as given
by the Ansatz (\ref{b_ansatz}). Bottom figure 2
shows that $\sum_{k} Y_{q,k}$ (which roughly 
measures the fermion pairs spectral weight) 
changes a lot for $q \sim 2k^{*}$. Again, 
if we recall that for a decreasing temperature 
$k_{F} \rightarrow k^{*}$ we can conclude 
that there is an increasing probability of 
finding more and more fermion pairs, even though 
our system is not in the superconducting phase. 
This is one of possible ways for observing 
the precursor effect. 

\begin{figure}
\centerline{\epsfxsize=8cm \epsfbox{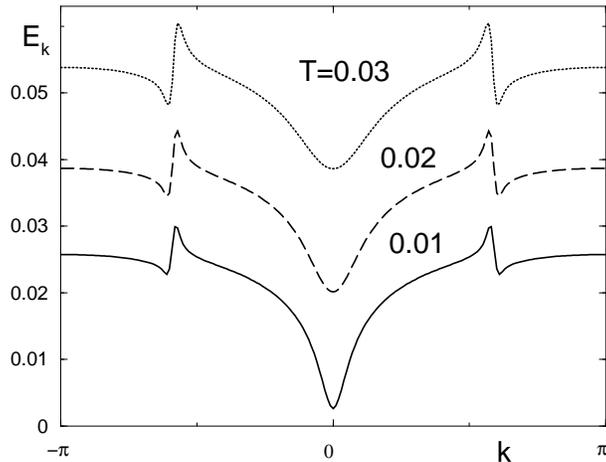}}
\caption{Effective boson dispersion $E_{k}(l=\infty)$ 
as measured from $2\mu$ level for the same set of temperatures 
as in figure 1. Recall that initially (for $l=0$ bosons
were localized.}
\end{figure}

In conclusion we studied evolution of fermion
and boson operators in the BF model. During a 
continuous transformation which eliminated the 
exchange interaction we observe transformation 
of the initial fermion and boson particles 
into the composite objects. In particular 
we find that fermion (boson) spectral weights 
at $k_{F}$ ($2k_{F}$) are reduced when temperature 
decreases. This effect is in an agreement with
the recent interpretation of the ARPES measurements
for underdoped Bi2212 cuprates. To clarify a
problem of the quasiparticles life time
we need a more detailed study of the correlation 
functions. Such an analysis is currently under 
investigation and the results will be published 
elsewhere.


\begin{thebibliography}{00}
\bibitem{Timusk}
         T.~Timusk and B.~Stratt, Rep.~Prog.~Phys.\
         {\bf 62}, 61 (1999).
\bibitem{BFM}
         J.~Ranninger and S.~Robaszkiewicz, Physica
	 {\bf B 135}, 468 (1985).
\bibitem{Robin}
         J. Ranninger, J.M. Robin and M. Eschrig,
         Phys. Rev. Lett. {\bf 74}, 4027 (1995); \\
         P. Devillard and J. Ranninger,
         Phys. Rev. Lett. {\bf 84}, 5200 (2000).
\bibitem{Ren}
         H.C.~Ren, Physica {\bf C 303}, 115 (1998).	 
\bibitem{NCA}
         J.M. Robin, A. Romano and J. Ranninger,
         Phys. Rev. Lett. {\bf 81}, 2756 (1998).
\bibitem{Domanski} T.~Doma\'nski and J.~Ranninger, Phys.~Rev.\
        {\bf B 63}, 134505 (2001).
\bibitem{Campuzano}
         M.R.~Norman, A.~Kami\'nski, J.~Mesot and J.C.~Campuzano,
         Phys.~Rev.\ {\bf B  63},140508 (2001).
\bibitem{spin-boson} S.~Kehrein and A.~Mielke, Ann.~Phys.\
         (Leipzig) {\bf 6}, 90 (1997).
\bibitem{Ragawitz} M.~Ragawitz and F.~Wegner, Eur.~Phys.~J. B
         {\bf 8}, 9 (1999).
\bibitem{Hofstetter} W.~Hofstetter and S.~Kehrein, Phys.~Rev.\
         {\bf B 63}, 140402 (2001).
\end{thebibliography}
\end{document}